\documentclass[conference]{IEEEtran}
\IEEEoverridecommandlockouts

\usepackage{cite}
\usepackage{amsmath,amssymb,amsfonts}
\usepackage{algorithmic}
\usepackage{graphicx}
\usepackage{textcomp}
\usepackage{xcolor}
\usepackage{hyperref}

\usepackage{amsmath,graphicx}
\usepackage{hyperref}
\usepackage{url}
\usepackage[usenames,dvipsnames]{pstricks}
\usepackage{epsfig}
\usepackage{pst-grad} 
\usepackage{pst-plot} 
\usepackage{amsmath,epsfig}
\usepackage{amssymb}
\usepackage{psfrag}
\usepackage{color}
\usepackage{pstricks}
\usepackage{cite}
\usepackage{algorithm}
\usepackage{epsfig}
\usepackage{amsthm}
\usepackage{amssymb}
\usepackage{psfrag}
\usepackage{pstricks}
\usepackage{float}
\usepackage{stfloats}
\usepackage{wrapfig}
\usepackage{graphicx}
\usepackage{amsmath}
\usepackage{amsfonts}
\usepackage{amssymb}
\usepackage{blindtext}
\usepackage{array}
\usepackage{arydshln}
\usepackage{pstricks,pst-plot}
\usepackage{pst-bezier}
\usepackage{pst-math}
\usepackage{algorithm,amsmath}
\usepackage{float}
\usepackage[mathscr]{eucal}
\usepackage{enumerate}

\usepackage{graphicx,psfrag}

\def\bp{{\bf p}}

\def\bW{{\bf W}}

\def\bh{{\bf h}}

\def\bH{{\bf H}}
\def\bP{{\bf P}}
\def\bS{{\bf S}}
\def\bQ{{\bf Q}}

\def\bB{{\bf B}}

\def\bB{{\bf B}}

\def\bs{{\bf s}}

\def\balpha{{\boldsymbol \alpha}}

\def\bbeta{{\boldsymbol \beta}}

\def\b0{{\bf 0}}

\DeclareMathAlphabet\mathbfcal{OMS}{cmsy}{b}{n}
\DeclareMathAlphabet\mathbfbb{OMS}{cmsy}{b}{n}

\def\BibTeX{{\rm B\kern-.05em{\sc i\kern-.025em b}\kern-.08em
    T\kern-.1667em\lower.7ex\hbox{E}\kern-.125emX}}
\begin{document}

\title{Graph Attention Network for Optimal User Association in Wireless Networks}


\author{\IEEEauthorblockN{Javad Mirzaei}
\IEEEauthorblockA{\textit{Chief Technology Office} \\
\textit{Dell Technologies Inc.}\\
Canada\\
javad.mirzaei@dell.com}
\and
\IEEEauthorblockN{Jeebak Mitra}
\IEEEauthorblockA{\textit{Chief Technology Office} \\
\textit{Dell Technologies Inc.}\\
Canada\\
jeebak.mitra@dell.com}
\and
\IEEEauthorblockN{Gwenael Poitau}
\IEEEauthorblockA{\textit{Chief Technology Office} \\
\textit{Dell Technologies Inc.}\\
Canada\\
gwenael.poitau@dell.com}
}

\maketitle

\begin{abstract}
With increased 5G deployments, network densification is higher than ever to support the exponentially high throughput requirements. However, this has meant a significant increase in energy consumption, leading to higher operational expenditure (OpEx) for network operators creating an acute need for improvements in network energy savings (NES). A key determinant of operational efficacy in cellular networks is the user association (UA) policy, as it affects critical aspects like spectral efficiency, load balancing etc. and therefore impacts the overall energy consumption of the network directly. Furthermore, with cellular network topologies lending themselves well to graphical abstractions, use of graphs in network optimization has gained significant prominence. In this work, we propose and analyze a graphical abstraction based optimization for UA in cellular networks to improve NES by determining when energy saving features like cell switch off can be activated. A comparison with legacy approaches establishes the superiority of the proposed approach.
\end{abstract}


\section{Introduction}
\label{sec:intro}

With the fifth generation (5G) and beyond 5G (B5G) roll-out, various use cases have been enabled that go  beyond just providing connectivity for mobile devices. In particular, they can be categorized into enhanced Mobile Broadband (eMBB) for high data rate, Ultra-reliable low latency (URLLC) for low latency and highly reliable use cases and machine-to-machine Type Communication (mMTC) for the internet of things (IoT) devices. While enabling these features is widely anticipated, each of these use cases place stringent demands on the network resources along with a commensurate increase in complexity of network design and operation. Network operators are challenged not only by the management of such complex mobile environments but also by an aspect that has received very little attention in previous generations: network energy consumption. 

Recently, increased energy consumption of wireless networks due to greater BS deployment density  compared to previous generation of networks has gained widescale attention. In fact, the radio access network (RAN) is deemed to be responsible for greater than $70\%$ of network's total power consumption \cite{8794573}. While legacy networks were forced to maintain BSs in a constant ON state regardless of the traffic level, leading to reduced energy efficiencies, current 3GPP standards have had a significant focus on developing energy saving features (ESFs) for RAN elements. Moreover, it is well-known that traffic loads of cellular networks exhibit spatio-temporal variation, which essentially implies the BS doesn't need to be configured for full-power operation of the radio at all times. Changes in user location and behaviour combined with temporal dynamics of the network imply that a user may not always be connected to the network in a way that optimizes energy usage. Load-balancing the network to maximize network energy savings (NES) is therefore required to ensure that UEs are connected to the BS that leads to the least amount of energy consumption. Such an action may be actuated either based on a periodic trigger or due to certain BSs within a cluster reaching a critical power consumption threshold determined by the network operator. Identifying such issues and re-distributing the UEs per a minimal network energy consumption criterion is a non-trivial task as it can have a cascading effect through the network and thus needs sophisticated solutions that are hard to derive analytically given the dynamics of the cellular environment. 

While modeling wireless networks have been pursued in various ways for operational optimization,  graph neural networks (GNNs) have only recently been used to that end \cite{9979700,  9046288, 9072356, 9252917, 9618652}. They allow graph-structured data to be processed effectively, enabled by global parameterizations, rotational and shift invariance \cite{9979700}, which give rise to their generalizability and scalability over different network topology. They exploit the underlying graph-structure of data as well as contextual information that facilitate the optimization of operational aspects of wireless networks \cite{9618652}.

In this paper, we aim to minimize the overall network energy consumption to improve NES, by jointly minimizing the fixed and traffic dependent power consumption. The former is done by employing cell switch off (CSO) when feasible while the latter is performed by optimally associating the UEs to the best BS. To do so, we propose a GNN-based technique with attention mechanism for user association in cellular networks to improve the NES.  The wireless network is first represented as a homogenous graph, whereby the UEs are represented as the nodes of the graph and the edges denote the connection between the UEs and BS. Leveraging the underlying graph structure of the network, GNN determines when energy saving features can be activated. These features include re-associating the UEs to the neighbouring BSs, thereby managing physical resource block (PRB) utilization per BS, as well as the initiating cell switch off. Compared with a legacy user association techniques, such as RSRP and a genie-aided per-PRB SINR metric, the proposed GNN-based technique shows significant improvement in NES.

\section{System Model} \label{sec:sys_model}

Given that network energy consumption increases in proportion to the throughput requirements of the UEs, this section provides a brief description of the scenarios that we consider in order to exhibit the nuances of the problem itself and the capabilities of a GNN-based approach to effectively address them. We consider a cellular network with $N$ cells that are adjacent to each other as shown exemplarily in Fig.\ref{Net_GNN}. Each cell site is equipped with the $N_{Tx}$ transmit antennas at a height of $h_{Tx}$. In the region of coverage (RoC) of this cluster of  $N$ cells, $K$ different UEs are uniformly distributed with the height of $h_{UE}$  and $N_{Rx}$ received antennas at each UE location. The UEs are able to connect to one or more of the BSs that are part of the cluster and their connectivity is jointly optimized to maximize NES through the UA policy. Data and control information flows between each UE and BS over $N_{sc}$ OFDM subcarriers, with the carrier frequency of $f_{c}$, $W_{sc}$ subcarrier spacing, and bandwidth $W$ in a 5G NR compliant format.

%
%

\subsection{Power Consumption  Model}
\label{subsec:PC}
Modeling the power consumption of the BS is an important first step towards developing methods to optimize it. It is well-known that each BS has a fixed portion of power consumption that depends on the type of deployed BS e.g. macro, small-cell etc, and a traffic load dependent portion that varies per UE demands. In particular, the overall power consumption of the $n^{\rm th}$ BS comprising of the traffic dependent and fixed portions is given by $P^n = P^n_{fixed} + P^n_{Tr-dep}$,
where $P^n_{fixed}$ is the power consumption of the $n^{\rm th}$ BS when carrying no traffic, and $P^n_{Tr-dep}$ is the traffic dependent power consumption of the corresponding BS. $P^n_{Tr-dep}$ is further composed of contributions from the baseband and the radio components \cite{6056691}
\begin{align}\label{Power_BS_dep}
\vspace{-0.5cm}
  P^n_{Tr-dep} (\eta) =  P^n_{BB} (\eta) +  P^n_{Radio}(\eta),
\end{align}
where $P^n_{Radio}(\eta) \triangleq N_{\rm Tx}\left[\frac{1}{(1+\epsilon)\sigma_{\rm max} }\left(\eta + \epsilon P^n_{{\rm max}, PA} \right)\right]$, and $P^n_{BB} (\eta)$ denotes the power consumed by the baseband circuitry. In Eqn. \eqref{Power_BS_dep}, $\sigma_{\rm max}$ is the maximum efficiency of PA, when transmitting the maximum output power $P^n_{{\rm max}, PA}$ with $\eta$ being the load on the PA and $\epsilon$ is a PA design-dependent parameter \cite{8094316}. Using the PRB utilization ratio as a proxy for $\eta$, we define $\eta \triangleq \frac{N_{\rm PRB-used}}{N_{\rm PRB-total}}$, where $N_{\rm PRB-used}$ is the number of allocated PRBs, $N_{\rm PRB-total}$ is the total number of available PRBs, and $0 \leq \eta \leq 1$.

\section{Wireless Network Graphs} 
\label{Sec:graph_rep}

\begin{figure}[t]
    \centering
    \begin{minipage}[b]{1\linewidth}
  \centering
 \includegraphics[width=.75\linewidth]{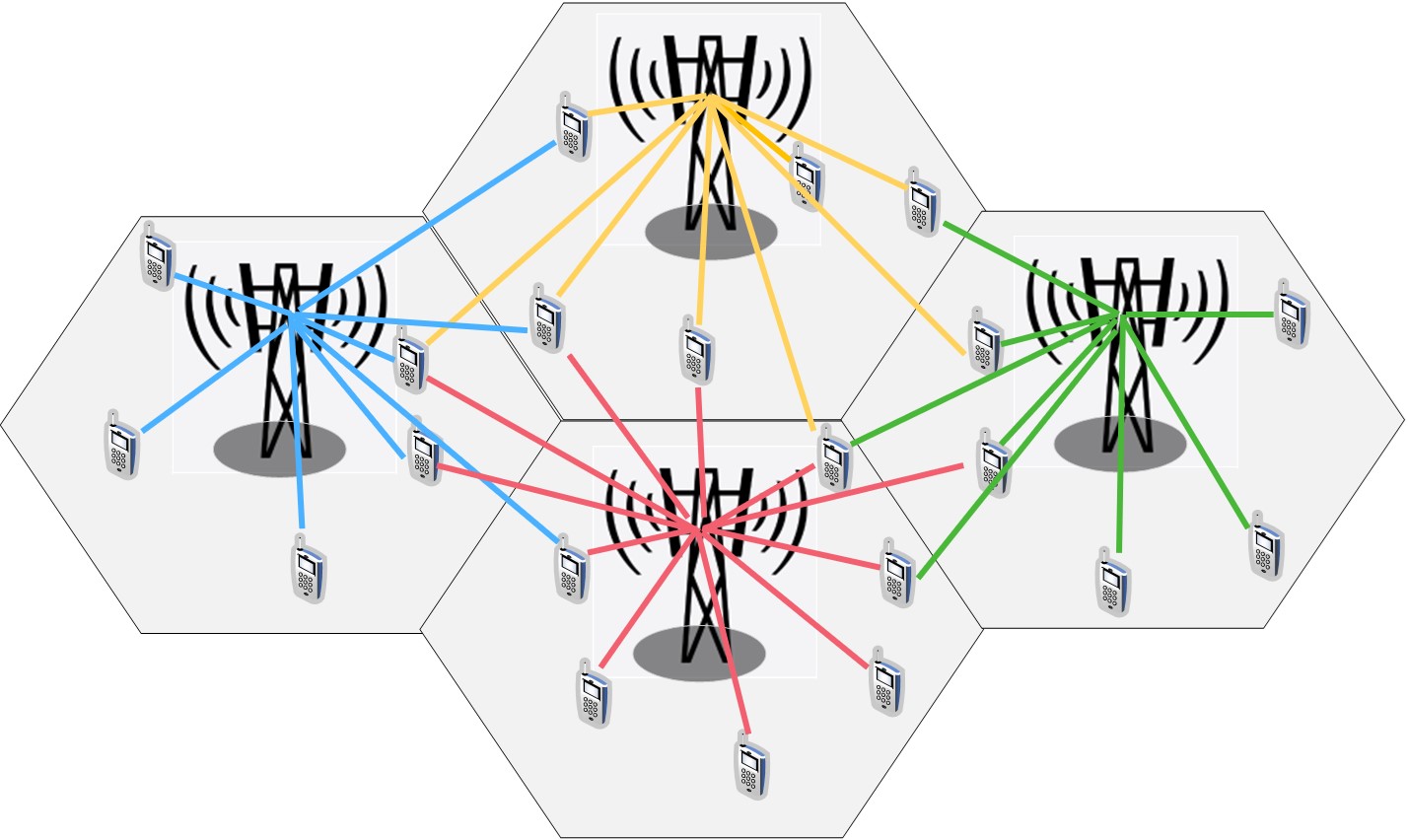}
	\caption{Logical connectivity graph representation between cell sites (BSs) and UEs.}
    \vspace{-0.5cm}
	\label{Net_GNN}
\end{minipage}
\end{figure}

The radio units (RUs) that can influence the energy consumption of each other, through actions such as user migration involving handovers through a change in UA policy triggered by load-balancing or NES considerations, are considered to be part of a network cluster. Fig. \ref{Net_GNN} provides a logical representation of the network graph, where the RUs and UEs are the nodes of this graph and the UE-RU connectivity is represented by the edges of the graph. While the corresponding network graph can be either represented heterogenous or homogenous,  in this work, we consider a homogenous graph  where all nodes and edges are considered to be of the same type. Mathematically, graphs can be expressed as $\mathcal{G}=(\mathcal{V},\mathcal{E})$ with $\mathcal{V}$ and $\mathcal{E}$ being the sets of nodes and edges, respectively. Let $v_i \in \mathcal{V}$ be a node in the graph and $e_{ij}=(v_i,v_j )\in \mathcal{E}$ be an edge from node $v_i$ to node $v_j$, then $e_{ij}=1$ indicates that $v_i$ and  $v_j$ are connected. The graph representation of the wireless network can be constructed by letting $\mathcal{G}({\bf A})=(\mathcal{V},\mathcal{E},{\bf H}^{(0)} )$ denote an equivalent network graph, where ${\bf H}^{(0)} \in R^{K \times d}$ represents the features’ matrix associated with the $\mathcal{G}$, and $d$ represents the number of features for each node of the graph. Let ${\bf h}_k^{(0)}$ denote the $k^{\rm th}$ row in ${\bf H}^{(0)}$, representing the $d$-dimensional feature vector for the $k^{\rm th}$ node. To construct ${ \bf h}_k^{(0)}$, we use a set of physically measured KPIs. In particular, we define ${\bf h}_k^{(0)}\triangleq \left[ {\bf p}_k,  {\bf q}_k, {\bf r}_k\right]$,
where ${\bf p}_k \triangleq \left[ p_{k1}, p_{k2}, \cdots, p_{kN} \right]$, ${\bf q}_k \triangleq \left[ q_{k1}, q_{k2}, \cdots, q_{kN} \right]$ and ${\bf r}_k \triangleq \left[ r_{k1}, r_{k2}, \cdots, r_{kN} \right]$ are the PRB, distance, and SINR vectors of the $k^{\rm th}$, $\forall k$, UE, respectively, where $p_{kn}$, $q_{kn}$, and $r_{kn}$, $\forall n$, are the PRB requirement, distance and the measured SINR between the $k^{\rm th}$ UE and the $n^{\rm th}$ BS, respectively.

Assuming $\mathcal{G}$ to be a homogenous graph, and let ${\bf A} \in\left\{0,1\right\}^{K \times K}$ denote the adjacency matrix of $\mathcal{G}$, a non-zero $(i,j)^{\rm th}$ entry of ${\bf A}$ denoted by $e_{ij} \in \mathcal{E}$, represents the edge connecting the nodes $i,j \in \mathcal{V}$. The adjacency matrix ${\bf A}$, is defined based on the SINR measurements such that for each BS, all the UEs whose measured SINR is greater than $\gamma^{\rm th}$, are connected. Mathematically, we define $ \mathcal{U}_n \triangleq \left\{ (i,j) | r_{in}> \gamma^{\rm th}\, {\rm and} \,r_{jn}> \gamma^{\rm th} \right\}$, and $\mathcal{U} \triangleq \bigcup_{n=1}^N \mathcal{U}_n$, where $r_{in}$, $\forall n$, being the SINR measurement at the $i$-th UE from cell $n$. Therefore, 
\begin{align}\label{A_def}
\vspace{-0.4cm}
{\bf A}(i,j) \triangleq \begin{cases}
                1  &  (i,j) \in \mathcal{U}\\
                0  &  {\rm Otherwise}.
                \end{cases}
\vspace{-0.4cm}
\end{align}
The next section delves into the details on how GNNs are leveraged to achieve the NES employing a combination of CSO and improved UA policies.
\section{GNN for NES-Optimized User Association} 
\label{Sec:GNN_NES}

Most NES opportunities present themselves when network traffic is low in the coverage area of a cell. This provides an opportunity for cell sites with negligible traffic to be completely switched off resulting in significant energy savings \cite{9606569, 8794573, 8014292}. However, when a cell with a few UEs generating a small amount of traffic is switched off, those UEs need to be migrated to active neighboring cell(s) that can handle additional traffic from the cell being switched off. Note that while this may increase the energy consumption of the cells to which the UEs are migrated, since a cell site is switched off completely, it is likely to still lead to a net benefit from a global network energy consumption perspective, if the energy saved by switching off the cell is greater than the nominal increase experienced by the new anchor cell.

We define the cell association process, by denoting $\bs_k \in \left[0, 1\right]^{1 \times N}$, $k= 1, 2, \cdots, K$, the association of the $k^{\rm th}$ UE to all cells. The $n^{\rm th}$  entry of $\bs_k$ denotes the association probability of $k^{\rm th}$ UE to the BS in the $n^{\rm th}$ cell. Therefore, the entries of $\bs_k$ are summed to $1$, i.e., $|\bs_k|_1 =1$. Stacking $\bs_k$, $k=1,2, \cdots, K$, we define the cell association matrix, $\bS \in \left[0,1\right]^{K \times N}$, where $\bS(k,n)$ represents the probability of association of the $k^{\rm th}$  node (UE) to the BS in the $n^{\rm th}$ cell. Furthermore, by stacking $\bp_k$, $\forall k$, we construct $\bP \in \mathbb{R}^{K \times N}$, the UE-BS  PRB allocation matrix. This allows us to define $\hat{\bp} \triangleq \bS^T \bP$, a $1 \times N$ vector, whose $n^{\rm th}$ entry denotes the total PRBs needed to serve the UEs associated with the BS in $n^{\rm th}$ cell for a given cell association matrix $\bS$.

\subsection{NES-based Optimization Objective }
The overall objective is to minimize the total energy consumption (EC) of the network. Additionally, we aim to make sure that the network traffic is well supported within the EC constraint. Expressing this mathematically, the total power consumption of the network is given by $P_{NW}= \sum_{n=1}^{N} P^n$, where $P^n$ is descirbed in Sec.\ref{subsec:PC}. Note that the fixed power consumption component of an active BS cannot be reduced by influencing network behaviour. However, when a BS can be switched off, a significant amount of saving is achieved if no serious degradation to network KPIs is anticipated. Such decisions are made as part of an overall objective of NES based UA and can be formalized as follows to minimize the average network energy consumption, 
\begin{align}\label{min_PNW}
\vspace{-0.4cm}
\min_{\mu}P_{NW}(\mu) = \sum_{n=1}^{N} P^n_{fixed} + P^n_{Tr-dep},
\vspace{-0.4cm}
\end{align}
where $\mu$ is the policy that leads to the least power consumption by user re-association with or without cell switch off. Additionally, denoting the aggregate throughput of the network cluster being optimized at the time $t$ by $L_t(\mu)$, a further constraint is put in order to ensure that the guaranteed bit rate (GBR) traffic doesn’t suffer as given by 
\begin{align}\label{C0}
\vspace{-0.4cm}
\min_{\mu} L_t(\mu) \geq L_{\rm min}, 
\vspace{-0.4cm}
\end{align}
where $L_{\rm min} = \sum_{n=1}^{N_{\rm GBR}} L_t^n$, and $N_{\rm GBR}$ denotes the total number of BSs that are carrying GBR traffic. Note that the optimization in \eqref{min_PNW}, given the constraint in \eqref{C0}, is an NP-hard problem. This typically requires an exhaustive search that is extremely computationally demanding due to the high dimensionality of the search space.

\subsection{Learning on GNN to achieve Improved NES}\label{GAT_def}

 Here, the ability of GNNs to be scalable is leveraged to not only address user association at a cluster level but also to take global view of the network when user migrations result in transfer to cells outside the cluster being considered. GNNs are constructed in a layered architecture where in each layer, the GNN updates the representation of each node as follows: 1) aggregating features from the node’s immediate neighbors only connected via the edges, 2) combining the node’s features with those aggregated from their neighbors via a \emph{permutation invariant} process. The update rule of the $l^{\rm th}$ layer at the node $u$ is given by \cite{veličković2018graph}
\begin{align}\label{beta_l}
\vspace{-0.3cm}
\bbeta_u^{(l)} &= \texttt{AGGREGATE}^{(l)}\left(\left\{ \bh_v^{(l-1)}, \forall v \in \mathcal{N}(u)\right\} \right),\\
\vspace{-0.3cm}
\bh_u^{(l)}&= \texttt{COMBINE}^{(l)}\left(\left\{ \bh_u^{(l-1)}, \bbeta_u^{(l)}\right\} \right),
\vspace{-0.3cm}
\end{align}
where $\bh_u^{(l)}$ denotes the feature vector aggregated by the node $u$ from its neighbors at the $l^{\rm th}$ layer, $\mathcal{N}(u)$ is the set of neighboring nodes to the node $u$, and $\bbeta_u^{(l)}$ represents the feature vectors of node $u$ at the $l^{\rm th}$ layer. $\texttt{AGGREGATE}^{(l)} (\cdot)$ and $\texttt{COMBINE}^{(l)}(\cdot)$ are used to denote the aggregation and combining functions of the $l^{\rm th}$ layer which can have various forms depending on the application. For example, a common variant of GraphSAGE \cite{GNN_Sage} performs an element-wise mean as $\texttt{AGGREGATE}^{(l)} (\cdot)$, followed by concatenation with $\bh_u^{(l-1)}$, a linear layer and a ReLU as $\texttt{COMBINE}^{(l)}(\cdot)$.

Many popular GNN architectures weigh all neighbors $v \in \mathcal{N}(u)$ with equal importance (e.g., mean or max-pooling as AGGREGATE) \cite{veličković2018graph} which is sometimes a limitation due to the relative importance of nodes. Graph attention network (GAT) \cite{veličković2018graph} addresses this by producing node representations computed through a learned weighted average of the representations of all neighbors. A scoring function $\rho : \mathbb{R}^d \times \mathbb{R}^d \rightarrow \mathbb{R}$ computes a score for every edge $e_{uv} \in \mathcal{E}$, indicating the importance of the features of the neighbor $u$ to the node $v$:
\begin{align}\label{rho}
\resizebox{.9\hsize}{!}{$\displaystyle \rho_{uv}^{(l)} \triangleq \texttt{LeakyReLU}_\delta \left( {\balpha^{(l)}}^T \cdot \left[\bW^{(l)}\bh_u^{(l-1)}\| \bW^{(l)}\bh_v^{(l-1)}  \right] \right)$},
\end{align}
where $\balpha^{(l)} \in \mathbb{R}^{2d_l'}$, $\bW^{(l)} \in  \mathbb{R}^{d_l' \times d_l}$ are learnable parameters, $\texttt{LeakyReLU}_\delta$ is a leaky ReLU non-linear function with negative slope $\delta$, and $(\cdot)^T$ is a transposition operation. These attention scores are normalized across all neighbors $u \in \mathcal{N}(v)$ using softmax, and the attention function is defined as 
\begin{align}\label{alpha_uv}
\alpha_{uv}^{l} \triangleq \texttt{Softmax}\left(\rho_{uv}^{(l)}\right) = \frac{\exp(\rho_{uv}^{(l)})}{\sum_{j \in \mathcal{N}(u)}\exp(\rho_{uj}^{(l)})}.
\end{align}
In our model, a GAT computes a weighted average of the transformed features of the neighbor nodes (followed by a nonlinear function $\sigma$) as the new representation of node $u$, using the normalized attention coefficients:
\begin{align}\label{h_ul}
\resizebox{.9\hsize}{!}{$\displaystyle \bh_u^{(l)} = \sigma \left(\alpha_{uu}^{l} \bW^{(l)} \bh_u^{(l-1)} + \sum_{v \in \mathcal{N}(u)} \alpha_{uv}^{l} \bW^{(l)} \bh_v^{(l-1)} \right)$},
\end{align}
where $h_u^{(l)}\in \mathbb{R}^{d_l}$ is the $d_l$-sized vector of embeddings generated by the $l^{\rm th}$ layer of GAT. Once $h_u^{(L)}$, $\forall u$ is obtained after passing through $L$ layers of GAT, they are passed through a fully connected neural network layer followed by a Softmax operation to find the cell association matrix $\bS$ i.e.,
\begin{align}\label{S_mtx}
\bS = \texttt{Softmax}\left( \sigma \left( \bH^{(L)}\bQ + \bB \right) \right),
\end{align}
where $\bQ \in \mathbb{R}^{d_L \times K} $ and $\bB \in \mathbb{R}^{K \times N}$ are the learnable weight and bias matrices, respectively.
\vspace{-0.2cm}

\subsection{Loss Function and GNN Training for NES }
Denoting by $\theta$ the set of parameters used to characterize the GNN, $P_{NW}$ can be reformulated to define the loss function as:
\begin{align}\label{l_theta}
\resizebox{.87\hsize}{!}{$\displaystyle \mathcal{L}(\theta) \triangleq P_{NW}^{\theta}(\mu) + \underbrace{\lambda_1 \left[K - {\rm Tr}\left(\bS(\theta) \bS^T(\theta)\right) \right]}_{T_1} + \underbrace{\lambda_2 |\hat{\bp}(\theta)|_2}_{T_2}$},
\end{align}
where $\lambda_1$ and $\lambda_2$ are the regularizing coefficients (i.e., hyper-parameters) used to control the training of GNN, and ${\rm Tr} (\cdot)$ is a trace operation. $P_{NW}^{\theta}(\mu)$ is the power consumption parameterized by $\theta$, that we aim to minimize. $T_1$ is the regularizing term that is meant to find the optimal UA. In particular, it ensures that each UE is associated with the BS of \emph{only one} cell. Since $|\bs_k|_1 =1$, and using the fact that $|\bs_k |_2 \leq |\bs_k |_1=1$, with equality only when \emph{one} entry of $\bs_k$ is $1$ and the rest of them to be $0$, we argue that ${\rm Tr}\left(\bS(\theta) \bS^T(\theta)\right)\leq K$.
The equality occurs when in each row of $\bS(\theta)$, only a single entry is $1$ while the rest are of entries being $0$. Note that with $T_1$, we intend to associate each UE to a BS in \emph{only one} cell via minimizing the gap between ${\rm Tr}\left(\bS(\theta) \bS^T(\theta)\right)$ and $K$. $T_2$  in \eqref{l_theta} is meant to control the distribution of PRBs (i.e., PRB utilization) across the BS in all cells, in a way that best minimizes $\mathcal{L}(\theta)$. As we show via the simulation results the effect of $T_2$ is highly depend of the operating bandwidth. We obtain $\theta$ by training the GNN using an unsupervised loss function given in \eqref{l_theta} via the back propagation technique. 

\section{SIMULATION RESULTS} 
\label{Sec:Sim}
This section provides the simulation results of the proposed GNN-based approach to NES-optimized UA.  

\subsection{Data Generation and Simulation Parameters}
The simulation scenario is composed of $N = 7$ cells with BS at fixed locations with an inter-site distance of $1$ km. The region of coverage is an area of size $3$ km $\times$ $3$ km with the UEs being uniformly distributed. Each cell site is equipped with the $N_{\rm Tx}=4$ transmit antennas with the height of $h_{\rm Tx}=25$ m.  The UEs are equipped with single antenna and $h_{\rm Tx} = 1.5$ m. The downlink communication between each UE and BS is 5G NR compliant and is based on clustered delay line (CDL) channel model in an Urban Macro (UMa) environment with $f_{c} =3.5$ GHz and $W_{sc}=30$ KHz. . We consider no UE mobility and therefore there is no Doppler shift in the channel. 


The cell sites are operated with a frequency reuse factor of $1/3$. As a benchmark, we compare the performance of the proposed GNN-based technique with: I) RSRP: UA is based on the generally-used wideband RSRP, II) Genie-aided sub-band SINR (GA-SubSINR): UA is performed based on the best sub-band SINR measured per-PRB basis to account for the frequency selectivity of the wireless channel.

\subsection{Network Architecture and Training}

The GNN is constructed with $2$ layers of GAT, and in the first layer, the GAT takes the network graph $\mathcal{G}$ with $d=21$ dimensional node features and maps the features into an embedding space of size $512$. After the first layer of GAT, we obtain the nodes’ representation $H^{(1)}$ in an embedding space. In the second layer of GAT, $H^{(1)}$ is mapped to another embedding space of size $512$, namely $H^{(2)}$. A linear transformation is applied to  $H^{(2)}$ with a bias, followed by a non-linear function $\sigma(\cdot)$, to map the embedded features to a vector of size $7$. These features are passed through the $\texttt{Softmax}(\cdot)$ operation to find $\bS$. We consider ReLU as an activation function $\sigma (\cdot)$ and $\delta = 0.2$. The entire implementation is done with PyTorch.

The size of data set is $10000$. To form the training and testing portions, we first collect the nodes’ features followed by scaling, normalizing and adding a self-loop for each node. A homogenous graph is then constructed for each network realization. The data set is then shuffled and split such that $80\%$ is used for training and $20\%$ for testing. ADAM optimizer is used to train the GNN with Learning rate $ = 5e^{-5}$, and $\beta  = (0.9, 0.999)$ in $5000$ epochs.

\subsection{NES Performance Gain vs Bandwidth}
\begin{figure}[!ht]
	\centering
	\includegraphics[width=1\linewidth]{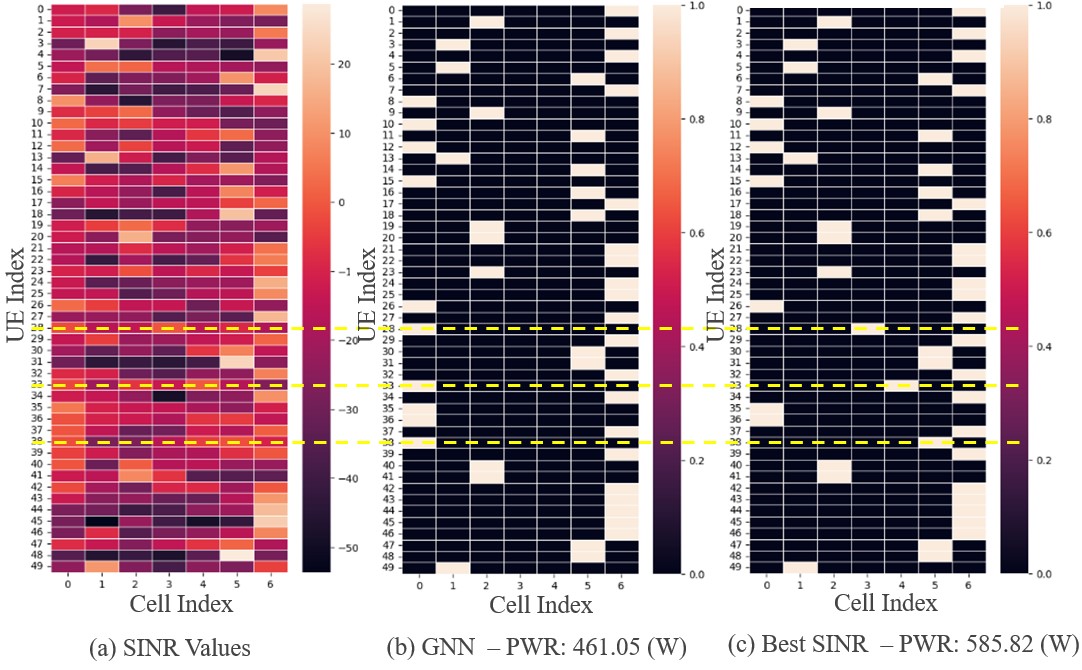}
\vspace{-.6cm}
	\caption{An example of user association for $50$ UEs across $7$ cells and $20$ (MHz) bandwidth: (a) SINR matrix, the values indicates the strength of SINR at each UE location for each BS. (b) The UE association based on the GNN technique. The $(k,n)$ entry indicates the association of $k^{\rm th}$ UE to the BS in the $n^{\rm th}$ cells. (c) The UE association based on the best SINR. }
	\label{Net_GNN1}
\end{figure}

\begin{figure}
	\centering
	\includegraphics[width=1\linewidth]{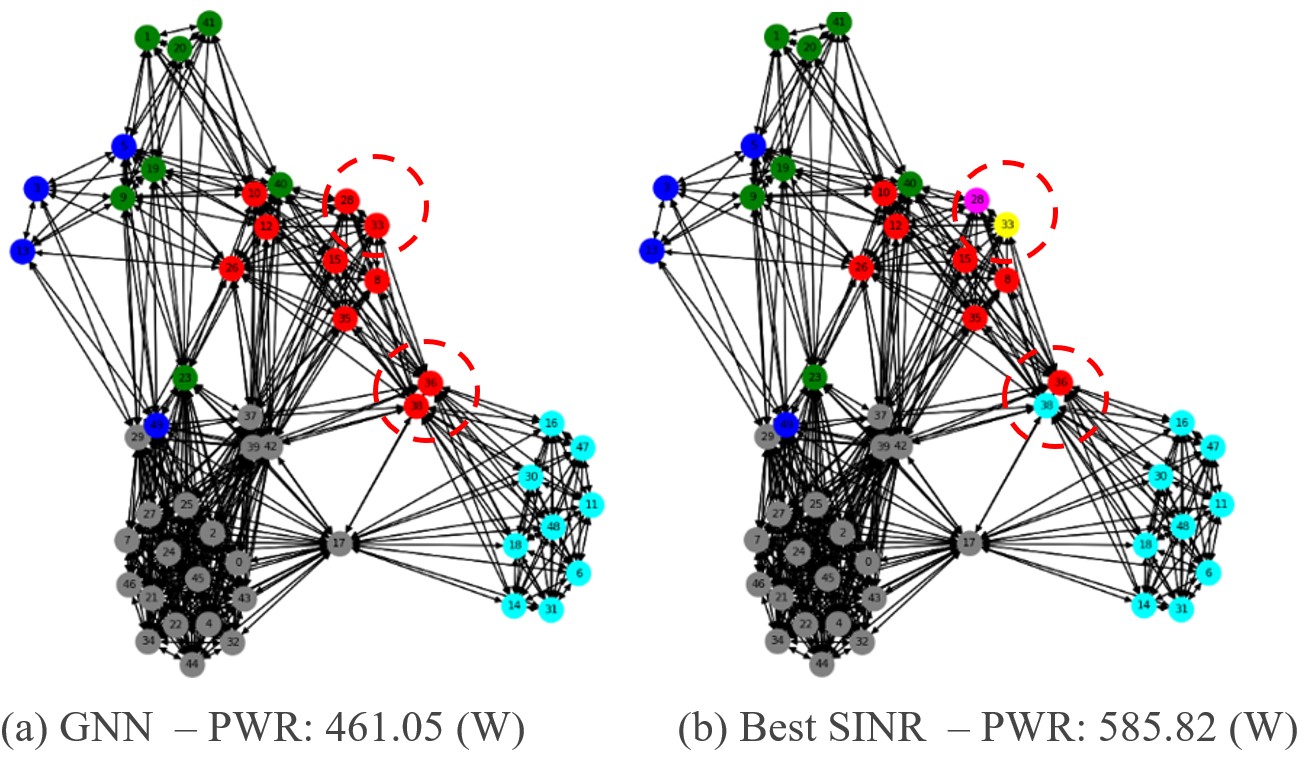}
\vspace{-.6cm}
	\caption{The graph representation of the UE association of the example in Fig. \ref{Net_GNN1}. (a) The graph corresponding to the GNN technique. (b) The graph corresponding to the best SINR. Each color represents the UEs connected to the BS in the same cell.}
\vspace{-.6cm}
	\label{Net_GNN2}
\end{figure}

\vspace{-0.1cm}
The heatmap in Fig. \ref{Net_GNN1} provides an example of the UE association using GNN and compared it with GA-SubSINR as a benchmark. This example is configured for $50$ UEs in the vicinity of a $7$-cell, and $20$ (MHz) of bandwidth. We observe that the connectivity in only few UEs are different namely the UE index of $28$, $33$, and $38$, which are highlighted in dash yellow horizontal lines. Such UE re-association has led to the cell site $3$ and $4$ to be completely shut down as their UEs are migrated to the neighboring cells. The corresponding graphical representation is provided in Fig. \ref{Net_GNN2}. It can be seen that while the connectivity remains largely the same the GNN based approach provides about a $21\%$ improvement in NES. This can witness significant benefits if we determine cell sites that can be switched off based on traffic load experienced. We will provide more detailed analysis on this later in this section. 


\begin{figure*}[ht]
    \centering
    \begin{minipage}[b]{.48\linewidth}
  \centering
  \centerline{\includegraphics[width=9.5cm]{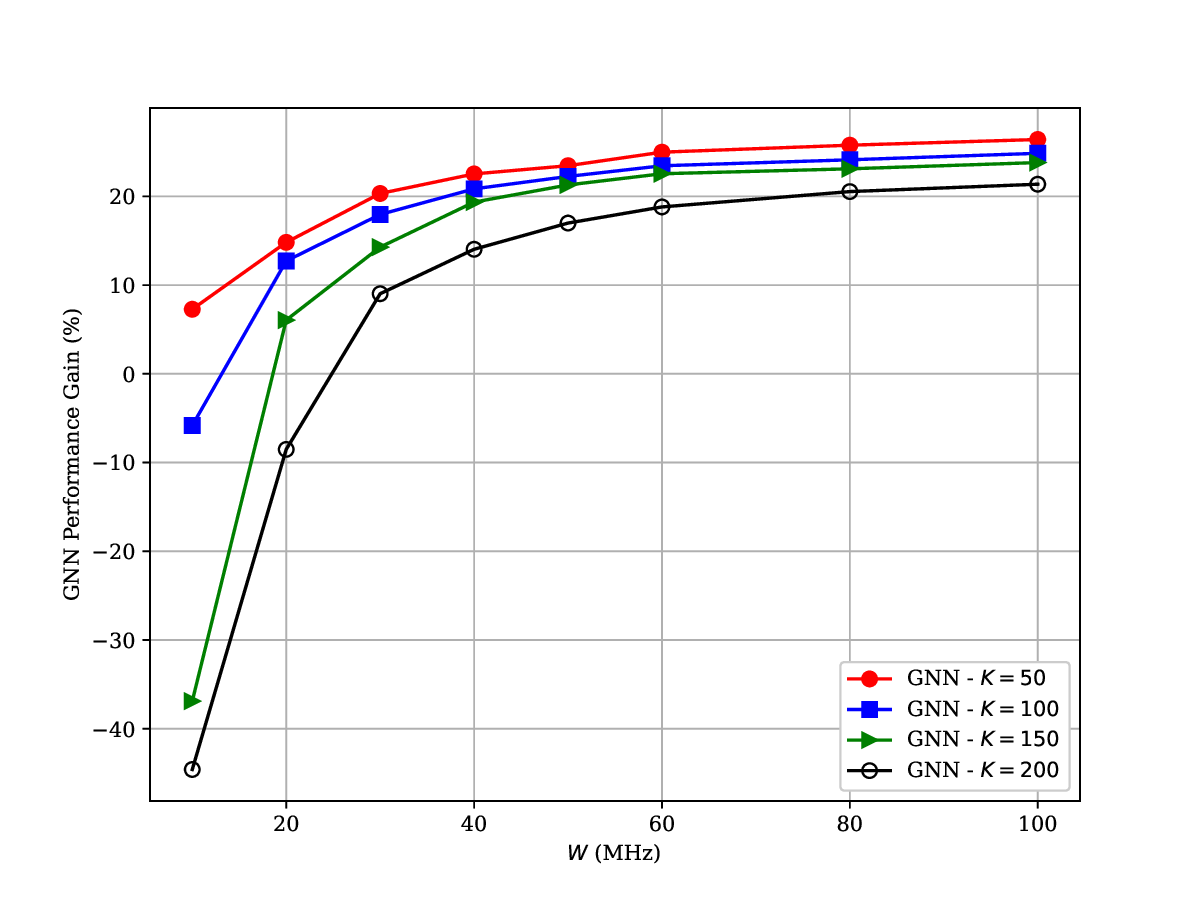}}
    \vspace{-0.4cm}
\caption{GNN performance gain ($\%$) vs $W$, Performance gain with respect to GA-SubSINR.}
\vspace{-0.4cm}
\label{PWR_vs_BW_SINR}
\end{minipage}
    \hfill
 \begin{minipage}[b]{.48\linewidth}
  \centering
  \centerline{\includegraphics[width=9.5cm]{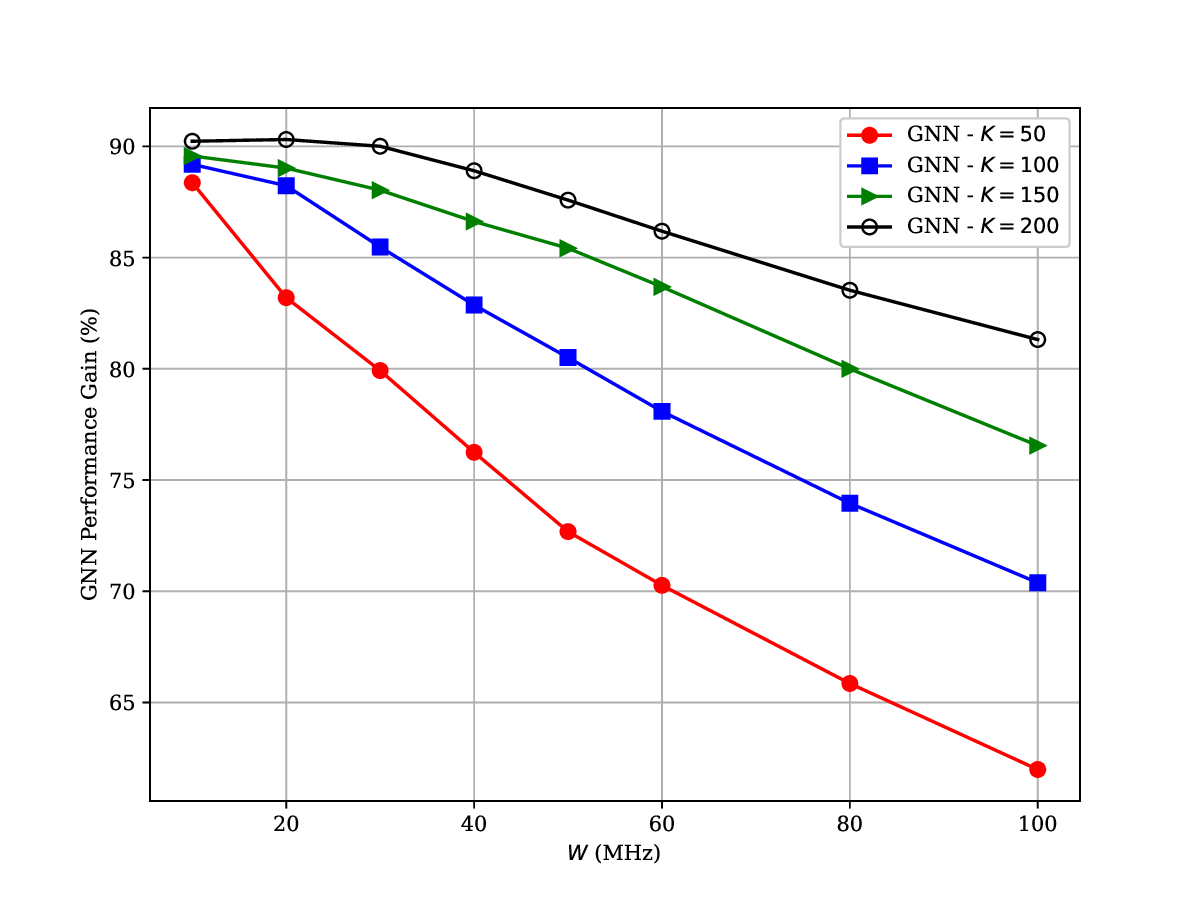}}
    \vspace{-0.4cm}
\caption{GNN performance gain ($\%$) vs $W$, Performance gain with respect to best RSRP.}
\vspace{-0.4cm}
\label{PWR_vs_BW_RSRP}
\end{minipage}
\end{figure*}

The GNN performance gain (in percentage) with respect to  GA-SubSINR and RSRP-based techniques are depicted in Fig. \ref{PWR_vs_BW_SINR} and  Fig. \ref{PWR_vs_BW_RSRP}, respectively, with increasing cell bandwidth. While the NES improvement compared to a legacy RSRP-based UA approach is fairly evident (ranging between $60-90\%$ according to Fig. \ref{PWR_vs_BW_RSRP}), the benefits are seen to diminish as more PRB resources become available with higher cell bandwidth. In Fig. \ref{PWR_vs_BW_SINR}, the GNN performance gain is in a more nominal range of $10-25\%$, which is still significant considering the GA-SubSINR approach has perfect channel information on a per sub-band basis. For a fixed bandwidth, the GNN-based technique performs better for small number of UEs rather large number of UEs. This can be explained by noting that for a fixed bandwidth (i.e., fixed PRB resources), the cell would have a relatively higher PRB utilization ratio (i.e., $\eta$) to serve the associated UEs. Such an increase in $\eta$ will lead to a relatively higher NES and therefore smaller performance gain. 

We also observe that, for a fixed number of UEs, the performance gain is not consistent across different cell bandwidths. In particular, for large bandwidth, the GNN-based technique provides the greatest performance gain. Due to the availability of relatively large PRB resources in large bandwidth, the overall PRB usage for each cell site remains relatively small which results in lower NES to service the UEs. Note however that, in extreme scenarios, where a large number of UEs are served within the small bandwidth, the GNN-based technique might fall short compared to what GA-SubSINR has to offer. As we show later via simulations, this can be addressed, through regularization terms, to keep the per-BS PRB utilization low. 
\subsection{Effect of Regularization}
\begin{figure*}[ht]
    \centering
    \begin{minipage}[b]{.48\linewidth}
  \centering
  \centerline{\includegraphics[width=9.5cm]{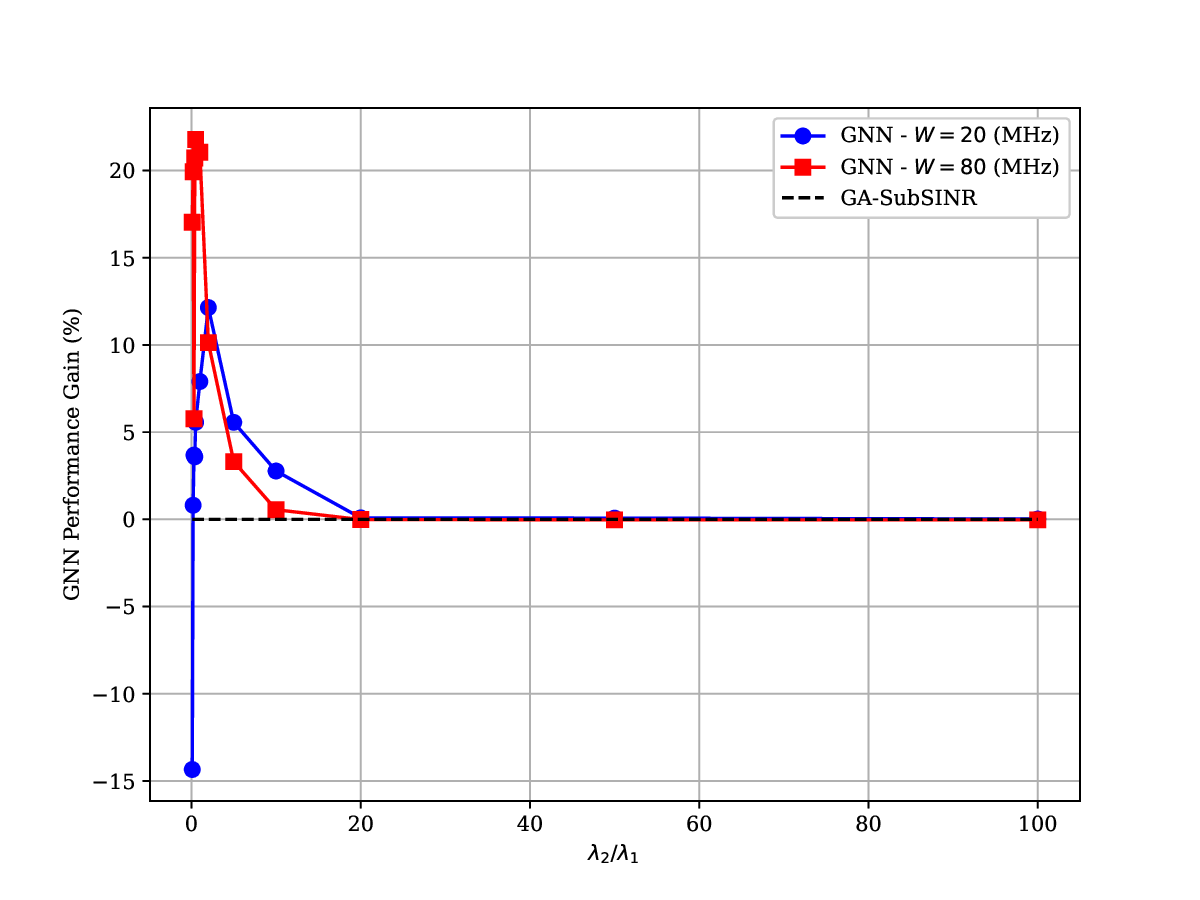}}
  \vspace{-0.3cm}
\caption{GNN performance gain ($\%$) with respect to GA-SubSINR vs $\lambda_2/\lambda_1$ for $W=[20, 80]$ (MHz) and $K=150$.}
\vspace{-0.5cm}
\label{Ratio150p}
\end{minipage}
    \hfill
 \begin{minipage}[b]{.48\linewidth}
  \centering
  \centerline{\includegraphics[width=9.5cm]{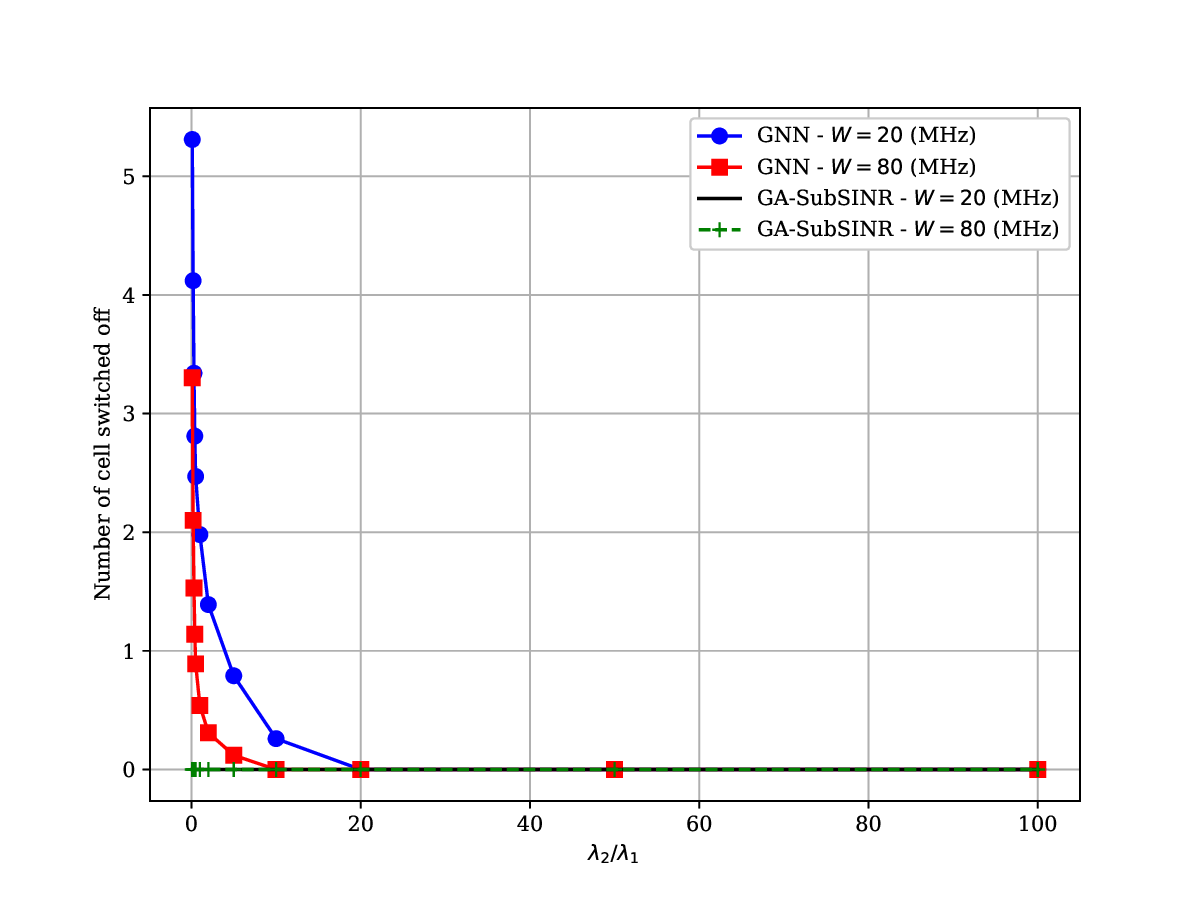}}
    \vspace{-0.3cm}
\caption{GNN-based Cell switched off rate vs $\lambda_2/\lambda_1$, for $W=[20, 80]$ (MHz) and $K=150$.}
\vspace{-0.5cm}
\label{CellOff150}
\end{minipage}
\end{figure*}

In Fig.  \ref{Ratio150p}, we explore the effect of $T_1$ and $T_2$ in training of GNN. In this figure, we plot the power consumption vs $\lambda_2/\lambda_1$ for $K=150$, $W = [20, 80]$ MHz and compare the prerformance of GNN-based technique to the GA-SubSINR as a benchmark. By varying $\lambda_2/\lambda_1$, we study the trade-off between the cell selection given by $T_1$ in $\mathcal{L}(\theta)$ and the network-level PRB distribution among the cell sites provided by $T_2$. For larger cell bandwidths, the GNN-based technique outperforms GA-SubSINR for all ranges of $\lambda_2/\lambda_1$. This is attributed to the availability of enough PRB resources to serve all UEs. For medium to large $\lambda_2/\lambda_1$ the GNN-based technique and GA-SubSINR perform the \emph{same} for both small and large bandwidth range. Similarly for very small $\lambda_2/\lambda_1$, GNN perform slightly better than GA-SubSINR. We note that, when $\lambda_2/\lambda_1= 0$, the GNN falls short compared to GA-SubSINR for $W = 20$ (MHz). This can be explained by noting that the term $T_2$ is ignored, the GNN may associate the UEs to BS irrespective of the PRB distributions among the cell sites, which may overwhelm certain BSs. At smaller bandwidth, this directly translates to higher power consumption. In contrast, for large bandwidth, there are enough PRB resources to accommodate all $150$ UEs without incurring further increase in power consumption. In both scenarios, the best performance gain is achieved when there is a balance between $T_1$ and $T_2$.

To further examine the effect of $T_1$ and $T_2$, we plot the number cell switch offs vs  $\lambda_2/\lambda_1$ in Fig. \ref{CellOff150}. We observe that for small $\lambda_2/\lambda_1$, where we tend to ignore $T_1$, a relatively large number of cell sites are switched off. In these cases, UEs are migrated to the neighbouring cells, incurring higher traffic loads. Compared to small bandwidth, larger cell bandwidths provide a better performance due to availability of enough PRBs to accommodate the same number of UEs. It is worth mentioning that, as  $\lambda_2/\lambda_1$ increases, the rate of cell switch off falls more rapidly for large bandwidth, due to the availability of more PRBs to serve the UEs in neighboring cells. For large $\lambda_2/\lambda_1$, $T_2$ becomes dominant, implying that PRB distribution among the cell sites are more even. Similar to GA-SubSINR, this leads to no cell switch off. 

\section{Conclusion} \label{Sec:conclusions} 
In this work we propose the use of GNNs to optimize user association from a NES perspective whereby the wireless network elements are modeled as nodes of a graph with the edges depicting connectivity of UEs with various base stations in its RF neighborhood. We formulate an optimization objective to improve the NES while maintaining a load-balanced network that prevents creation of network hotspots and automatically determines cell sites that may be switched off to transition the network to a more NES-optimal state. We also describe the architecture of our GNN-based approach along with the simulation scenarios constructed to do a robust evaluation. Comparisons with a legacy UA approach using RSRP shows significant improvement utilizing a GNN-based NES approach while it is shown to be competitive with a genie-aided approach where a per-PRB SINR metric is deemed available as an idealistic case. In future work, we will analayze the impact on QoS of the proposed NES policies and derive approaches that minimize impact. 

\bibliographystyle{IEEEbib}
\bibliography{strings,refs, references}

\end{document}